\documentstyle[12pt,epsfig]{article}
\def\be{\begin{equation}}
\def\ee{\end{equation}}
\def\bc{\begin{center}}
\def\ec{\end{center}}
\def\bea{\begin{eqnarray}}
\def\eea{\end{eqnarray}}

\def\fb{{\; \rm fb}}

\def\gev{{\rm \; GeV}}

\def\gt{\tilde{G}}

\def\mgg{M_{\gamma \gamma}}
\def\mgz{M_{\gamma Z}}

\def\mzz{M_{ZZ}}

\def\ov{\overline}
\def\pb{{\; \rm pb}}

\def\simlt{\stackrel{<}{{}_\sim}}

\def\tev{{\rm \; TeV}}

\catcode`@=11
\def\marginnote#1{}
\newcount\hour
\newcount\minute
\newtoks\amorpm
\hour=\time\divide\hour by60
\minute=\time{\multiply\hour by60 \global\advance\minute by-\hour}
\edef\standardtime{{\ifnum\hour<12 \global\amorpm={am}%
        \else\global\amorpm={pm}\advance\hour by-12 \fi
        \ifnum\hour=0 \hour=12 \fi
        \number\hour:\ifnum\minute<10 0\fi\number\minute\the\amorpm}}
\edef\militarytime{\number\hour:\ifnum\minute<10 0\fi\number\minute}
\def\draftlabel#1{{\@bsphack\if@filesw {\let\thepage\relax
   \xdef\@gtempa{\write\@auxout{\string
      \newlabel{#1}{{\@currentlabel}{\thepage}}}}}\@gtempa
   \if@nobreak \ifvmode\nobreak\fi\fi\fi\@esphack}
        \gdef\@eqnlabel{#1}}
\def\@eqnlabel{}
\def\@vacuum{}
\def\draftmarginnote#1{\marginpar{\raggedright\scriptsize\tt#1}}
\def\draft{\oddsidemargin 0.0truein
        \def\@oddfoot{\sl preliminary draft \hfil
        \rm\thepage\hfil\sl\today\quad\militarytime}
        \let\@evenfoot\@oddfoot \overfullrule 3pt
        \let\label=\draftlabel
        \let\marginnote=\draftmarginnote
   \def\@eqnnum{(\theequation)\rlap{\kern\marginparsep\tt\@eqnlabel}%
\global\let\@eqnlabel\@vacuum}  }
\catcode`@=12
\begin{document}
\begin{titlepage}
\vspace*{-1cm}
\phantom{hep-ph/0005076} 
\hfill{DFPD-00/TH/17}
\vskip 2.0cm
\begin{center}
{\Large\bf Signatures of massive sgoldstinos at hadron colliders}
\end{center}
\vskip 1.5  cm
\begin{center}
{\large Elena 
Perazzi}\footnote{e-mail address: perazzi@pd.infn.it}
\\
\vskip .1cm
Dipartimento di Fisica, Universit\`a di Padova, I-35131 Padua, Italy
\\
\vskip .2cm
{\large Giovanni Ridolfi}\footnote{e-mail address: ridolfi@ge.infn.it}
\\
\vskip .1cm
INFN, Sezione di Genova, I-16146 Genoa, Italy
\\
\vskip .2cm
and
\\
\vskip .2cm
{\large Fabio
Zwirner}\footnote{e-mail address: zwirner@pd.infn.it}
\\
\vskip .1cm
INFN, Sezione di Padova, I-35131 Padua, Italy
\end{center}
\vskip 2.5cm
\begin{abstract}
\noindent
In supersymmetric extensions of the Standard Model with a very
light gravitino, the effective theory at the weak scale should 
contain not only the goldstino $\gt$, but also its supersymmetric 
partners, the sgoldstinos. In the simplest case, the goldstino 
is a gauge-singlet and its superpartners are two neutral spin--0 
particles, $S$ and $P$. We study possible signals of massive 
sgoldstinos at hadron colliders, focusing on those that are most 
relevant for the Tevatron. We show that inclusive production of
sgoldstinos, followed by their decay into two photons, can lead 
to observable signals or to stringent combined bounds on the 
gravitino and sgoldstino masses. Sgoldstino decays into two 
gluon jets may provide a useful complementary signature.
\end{abstract}
\end{titlepage}
\setcounter{footnote}{0}
\vskip2truecm
\bc
{\bf 1. Introduction}
\ec
Sgoldstinos are the supersymmetric partners of the goldstino 
$\gt$, and, in supersymmetric models with a very light gravitino 
\cite{slgold}--\cite{bpz}, are expected to be part of the 
effective theory at the weak scale. In the simplest case, the R--odd 
goldstino is a gauge singlet, and its R--even superpartners are
two neutral spin--0 particles, $S$ (CP--even) and $P$ (CP--odd):
in the following, we will use the generic symbol $\phi$ to 
denote any of the two sgoldstino states. If the sgoldstino 
masses $m_{\phi}$ and the supersymmetry-breaking scale 
$\sqrt{F}$ are not too large,~\footnote{We remind the reader
that the supersymmetry--breaking scale $\sqrt{F}$ can be put in
one--to--one correspondence with the gravitino mass $m_{3/2}$,
via the relation $F = \sqrt{3} \, m_{3/2} M_P$, where $M_P
\equiv (8 \pi G_N)^{-1/2} \simeq 2.4 \times 10^{18} \gev$ is the
Planck mass.} sgoldstino production and decay may be detectable 
at the present collider energies.

Previous studies of sgoldstino phenomenology at colliders \cite{sgcol}
were performed under the very restrictive assumption of negligible
sgoldstino masses. In a recent paper \cite{prz}, we started a 
systematic investigation of the possible phenomenological signatures 
of massive sgoldstinos, concentrating on $e^+e^-$ colliders and 
in particular on LEP. We showed that the study of processes such
as $e^+ e^- \rightarrow \gamma \phi$, $Z \phi$ or $e^+ e^- \phi$, 
followed by $\phi$ decaying into two gluon jets, can explore virgin 
land in the parameter space of models with a superlight gravitino. 
This was confirmed by a subsequent experimental analysis \cite{delphi}, 
where (using not only the two-gluon decay mode, but also the two-photon 
one, as suggested by previous LEP searches \cite{opal}) new stringent 
combined bounds on the gravitino and sgoldstino masses were derived.

In the present work, we extend the study of the phenomenology of
massive sgoldstinos to the case of hadron colliders, with
emphasis on the Tevatron  ($p \ov{p}$; $\sqrt{S}=1.8 \tev$
and $L \sim 100 \pb^{-1}$ in run~I;  $\sqrt{S}=2 \tev$ and $L 
\sim 2 \fb^{-1}$ in run~II). Our study is organized as follows. 
In sect.~2, we complete the discussion of the sgoldstino 
effective couplings and branching ratios given in \cite{prz}, to 
encompass a range of sgoldstino masses extending above the $t \ov{t}$ 
threshold. In sect.~3, we discuss the mechanisms for sgoldstino production 
at hadron colliders, giving explicit formulae for the relevant partonic 
cross-sections and showing some representative numerical results.
In sect.~4, we review the resulting signals at the Tevatron and we 
attempt a first discussion of the discovery potential of the different 
channels, leaving a more detailed analysis to our experimental 
colleagues.  We show that inclusive production of sgoldstinos, followed 
by their decay into two photons, can lead to observable signals or to 
stringent combined bounds on the gravitino and sgoldstino masses. The
sgoldstino decay mode into two gluon jets, dominant over most of the
parameter space but plagued by large backgrounds, may provide a 
useful complementary signature when gluinos are much heavier than
the electroweak gauginos and higgsinos. Associated production with 
an electroweak gauge boson $(\gamma,W,Z)$ and/or other decay modes 
$(\gamma Z, ZZ, WW,\gt \gt, t \ov{t})$ do not lead in general to
an increased sensitivity.

To conclude this introduction, we would like to alert the reader on 
some simplifying assumptions underlying our analysis, and on the
analogies between sgoldstinos and other hypothetical spin--0 particles,
such as the neutral Higgs bosons of the Standard Model (SM) or
its Minimal Supersymmetric extension (MSSM). As in \cite{prz}, we 
will perform our study assuming that there is no mixing between 
sgoldstinos and Higgs bosons, and that R--odd MSSM particles and
Higgs bosons~\footnote{We remind the reader that, in models with
a very light gravitino, the MSSM upper bound \cite{erz} on the
mass of the lightest Higgs boson can be grossly violated 
\cite{bfz}.} are sufficiently heavy not to play a r\^ole in
sgoldstino production and decay. This should be taken as a 
benchmark case, which may be further generalized by including an
exhaustive treatment of the interplay between $SU(2) \times U(1)$ and 
supersymmetry breaking \cite{brignole}. Whenever possible, however,
we will compare the properties of our `pure' sgoldstinos with the
properties of the `pure' SM or MSSM Higgs bosons: in the case of 
non--negligible mixing, we may expect intermediate properties. In 
this respect, sgoldstinos represent a motivated and well-defined 
addition to a long list of other hypothetical spin--0 particles 
(`bosonic Higgs', `coloron', \ldots) that have been proposed with 
various theoretical motivations, and are often used in data analyses
to parametrize some of the searches for new physics. Since the 
indirect evidence for the SM (or the MSSM) as the correct and complete 
theory at the weak scale is far from conclusive \cite{bagger}, we 
believe that more exotic possibilities such as sgoldstinos still
deserve to be taken seriously, even on purely phenomenological 
grounds.  On the theoretical side, it is interesting to notice 
that sgoldstinos bear some similarities with the spin--0 fields 
arising from the metric (and antisymmetric tensors) of extra 
space-time dimensions in some `brane-world' scenarios \cite{add,rs}, 
whose collider phenomenology has been recently discussed \cite{radion}.
These similarities could be investigated more closely, leading perhaps 
to a more unified picture, if progress were made in the discussion of 
the breaking of supersymmetry and of the electroweak symmetry in this 
context.

\vspace{1cm}
\bc
{\bf 2. Sgoldstino effective couplings and branching ratios}
\ec

The general theoretical framework for the discussion of sgoldstino 
phenomenology was reviewed in \cite{prz}, to which we refer the reader. 
Sgoldstino effective interactions and branching ratios were also 
discussed in \cite{prz}, with emphasis on the sgoldstino mass range 
kinematically accessible at LEP, $m_{\phi} \simlt 200 \gev$. It was shown
that the sgoldstino couplings to gauge boson pairs can be parametrized
by the supersymmetry-breaking scale $\sqrt{F}$, by the gaugino masses 
$(M_3,M_2,M_1)$ and by a mass parameter $\mu_a$, associated with 
the charged higgsino and the antisymmetric combination of neutral 
higgsinos. To extend the discussion to higher sgoldstino masses, which 
may be phenomenologically relevant at the Tevatron (or eventually at 
the LHC), we need only a more detailed discussion of the $t \ov{t} 
\phi$ effective couplings and of $\phi \to t \ov{t}$ decays.

As discussed in \cite{bpz}, the Yukawa couplings of sgoldstinos to 
leptons and light quarks (for our purposes, all but the top quark)  
are expected to be suppressed by a factor $m_f/\sqrt{F}$, where 
$m_f$ is the fermion mass. We then expect these couplings to be 
important only for the top quark, and parametrize them as follows:
\be
\label{tth}
{\cal L}_{\phi t \ov{t}} = 
- {m_t A_S \over \sqrt{2} F} 
\left( S t t^c + {\rm h.c.} \right)
- {m_t A_P \over \sqrt{2} F} 
\left( i P t t^c + {\rm h.c.} \right) \, ,
\ee
where we used two-component spinors in the notation of \cite{prz},
and $A_S,A_P$ are free parameters with the dimension of a mass. Since 
$A_S+A_P=2 A_t$, where $\delta m^2 = m_t A_t$ is the off-diagonal
element in the mass matrix for the stop squarks, we expect $A_S$ and 
$A_P$ to be of the order of the other supersymmetry-breaking masses
in the strongly interacting sector of the theory. Notice that the 
$S t \ov{t}$ coupling is identical in form to the coupling of the 
SM Higgs, and can be obtained from the latter by performing the 
substitution
\be
\label{subs}
{g \over 2 m_W} \longrightarrow {A_S \over \sqrt{2} F} \, .
\ee
Similarly, the $P t \ov{t}$ coupling can be obtained from the coupling of 
the SM Higgs by inserting, in four-component notation, a $\gamma_5$ matrix 
in the fermion bilinear, and by performing the substitution
\be
\label{subp}
{g \over 2 m_W} \longrightarrow {i A_P \over \sqrt{2} F} \, .
\ee
Alternatively, the $P t \ov{t}$ coupling can be also obtained from the
$A t \ov{t}$ coupling of the MSSM by performing the substitution of 
eq.~(\ref{subp}) and by making the choice $\tan \beta = 1$. Notice also 
that, in general, we can have $A_S \ne A_P$. In the following, however, 
we will make the simplifying assumption $A_S = A_P \equiv A_t$, and, 
whenever needed for numerical examples, the representative choice $A_t 
= M_3$. From eq.~(\ref{tth}) we find
\be
\Gamma(S \to t \ov{t}) =
{3 m_t^2 A_S^2 m_S \over 16 \pi F^2}
\left( 1 - {4 m_t^2 \over m_S^2} \right)^{3/2} \, ,
\;\;\;\;\;
\Gamma(P \to t \ov{t}) =
{3 m_t^2 A_P^2 m_P \over 16 \pi F^2}
\left( 1 - {4 m_t^2 \over m_P^2} \right)^{1/2} \, .
\ee
Notice the different exponent for the CP-even and the CP-odd
state, as in the MSSM.  

We are now ready to extend the discussion of the sgoldstino
branching ratios to the mass range of interest at the Tevatron. 
We will focus our attention on the dependences on $m_{\phi}$ and 
$\sqrt{F}$, by making for the remaining parameters the two 
representative choices given in Table~1.
\begin{table}[ht]
\begin{center}
\begin{tabular}{|c|c|c|c|c|c|c|}
\hline
& $M_3$ & $M_2$ & $M_1$ & $\mu_a$ & $A_S$ & $A_P$ \\
\hline
(a) & 400 & 300 & 200 & 300 & 400 & 400 \\
\hline
(b) & 350 & 350 & 350 & 350 & 350 & 350 \\
\hline
\end{tabular}
\end{center}
\caption{Two representative choices for the mass parameters 
affecting the sgoldstino effective couplings. All masses are 
expressed in GeV.}
\end{table}
The corresponding chargino and neutralino masses (in GeV) are 
approximately $(220,380)$ and $(175,240,385)$ for case (a), $(270,
430)$ and $(260,350,440)$ for case (b). There is of course a fourth
neutralino, the symmetric higgsino combination, whose mass is
controlled by an independent parameter $\mu_s$, and does not 
affect the sgoldstino couplings. For most values of $\sqrt{F}$ 
to be considered in the following, the two parameter choices 
of Table~1 should be comfortably compatible with the present 
experimental limits on R-odd supersymmetric particles, coming 
from LEP and Tevatron searches. For sufficiently large values
of the sgoldstino masses, we should also consider sgoldstino 
decays into pairs of neutralinos, charginos or gluinos. However,
the relevant couplings are controlled not only by the parameters
of Table~1, but also by $\mu_s$, by the Higgs boson masses and by 
other parameters not related with the spectrum \cite{brignole}. In 
the following, we will always assume that the corresponding branching 
ratios can be safely neglected. In the region of sgoldstino masses 
considered in this paper, the only kinematically allowed decays could 
be those into the lightest neutralinos and charginos. For a wide range 
of the remaining parameters, these decays are sufficiently suppressed 
by the phase space and other factors, so that our discussion of the
dominant decay modes remains a good approximation. 

\begin{figure}[htbp]
\begin{center}
\epsfig{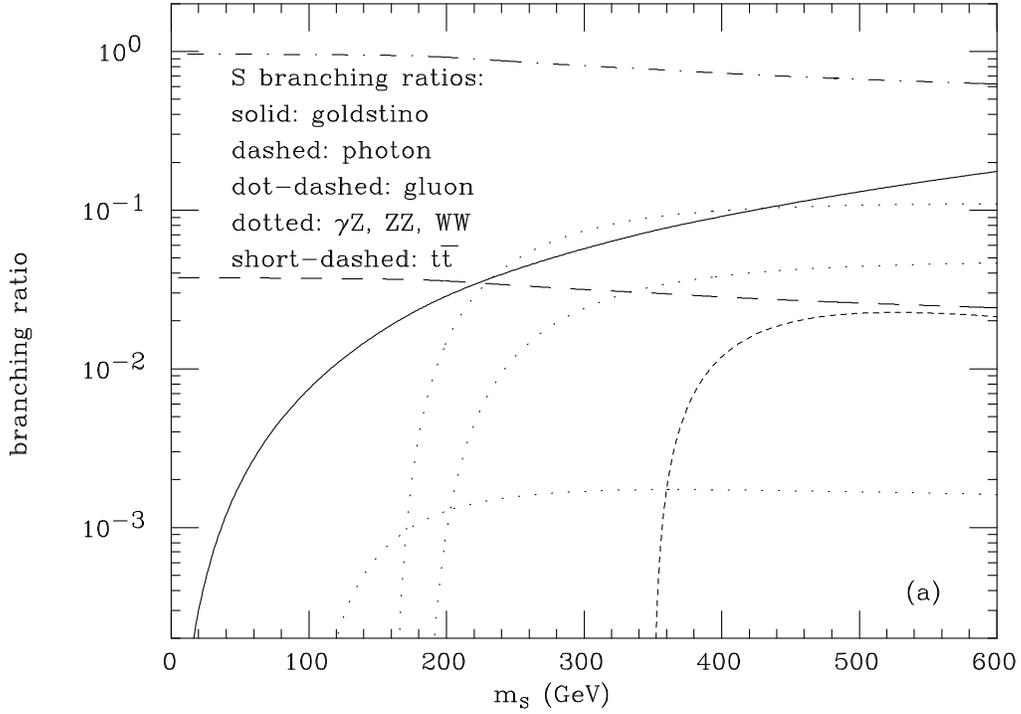}
\epsfig{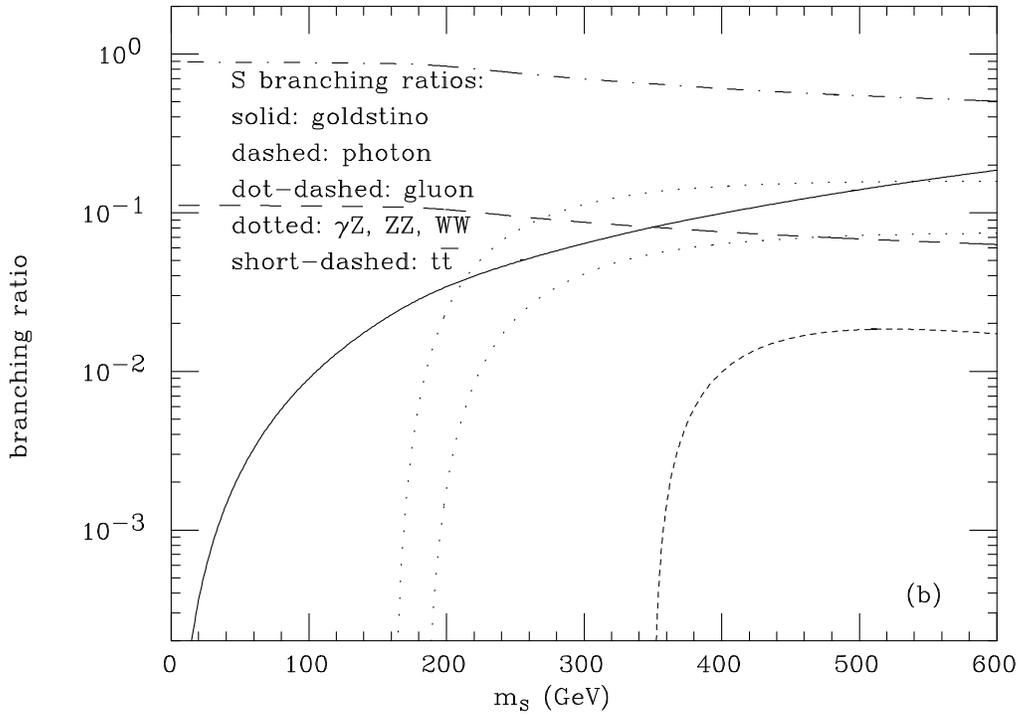}
\end{center}
\caption{{\it The $S$ branching ratios, as functions 
of $m_S$, for the two parameter choices of Table~1.}}
\label{brs}
\end{figure}

\begin{figure}[htbp]
\begin{center}
\epsfig{figure=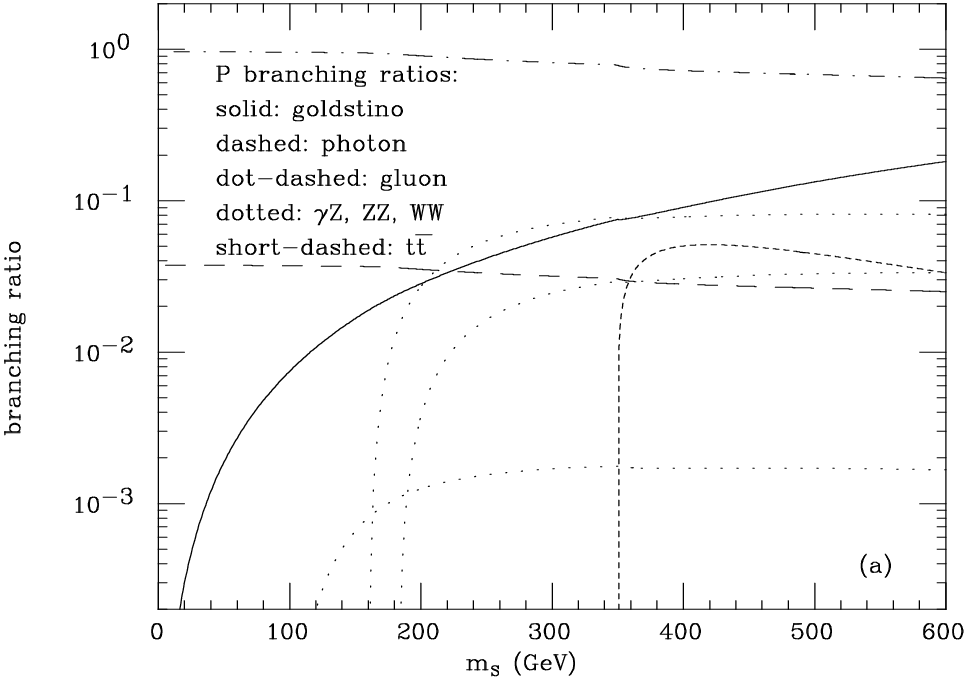,height=9.5cm,angle=0}
\epsfig{figure=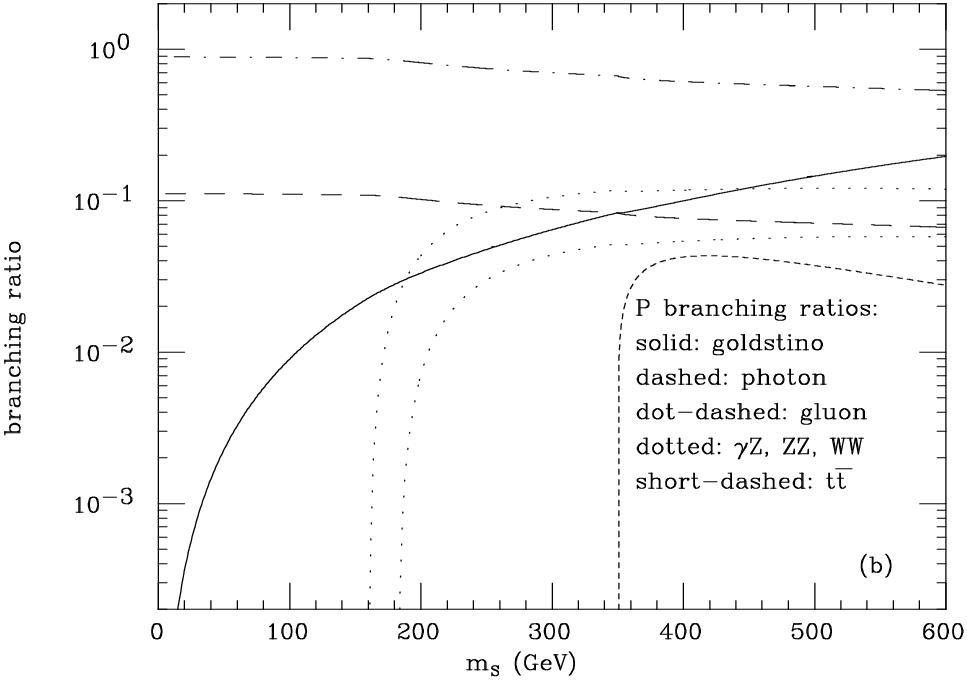,height=9.5cm,angle=0}
\end{center}
\caption{{\it The $P$ branching ratios, as functions 
of $m_P$, for the two parameter choices of Table~1.}}
\label{brp}
\end{figure}

Since all the partial widths for two--body decays are proportional to 
$F^{-2}$, the dependence on $F$ drops out of the discussion of the 
$\phi$ branching ratios. The latter are shown in Figs.~\ref{brs} and 
\ref{brp}, as functions of $m_{\phi}$, for the two cases of $S$
and $P$ and for the two parameter choices of Table~1.
We can see that the differences in the couplings of $S$ and $P$
to the massive weak bosons and to the top quark do not have an
important impact on the branching ratios. For both $S$ and $P$,
the most important decay mode is the one into gluons, with the 
one into goldstinos becoming competitive only for very large
sgoldstino masses. In the whole mass range up to 1~TeV, decays
into electroweak bosons are suppressed, at the level of $10 \%$
or less: as we will see, however, these modes could still be 
relevant for hadron collider searches, because of the much 
smaller backgrounds of the resulting signals.

\begin{figure}[htbp]
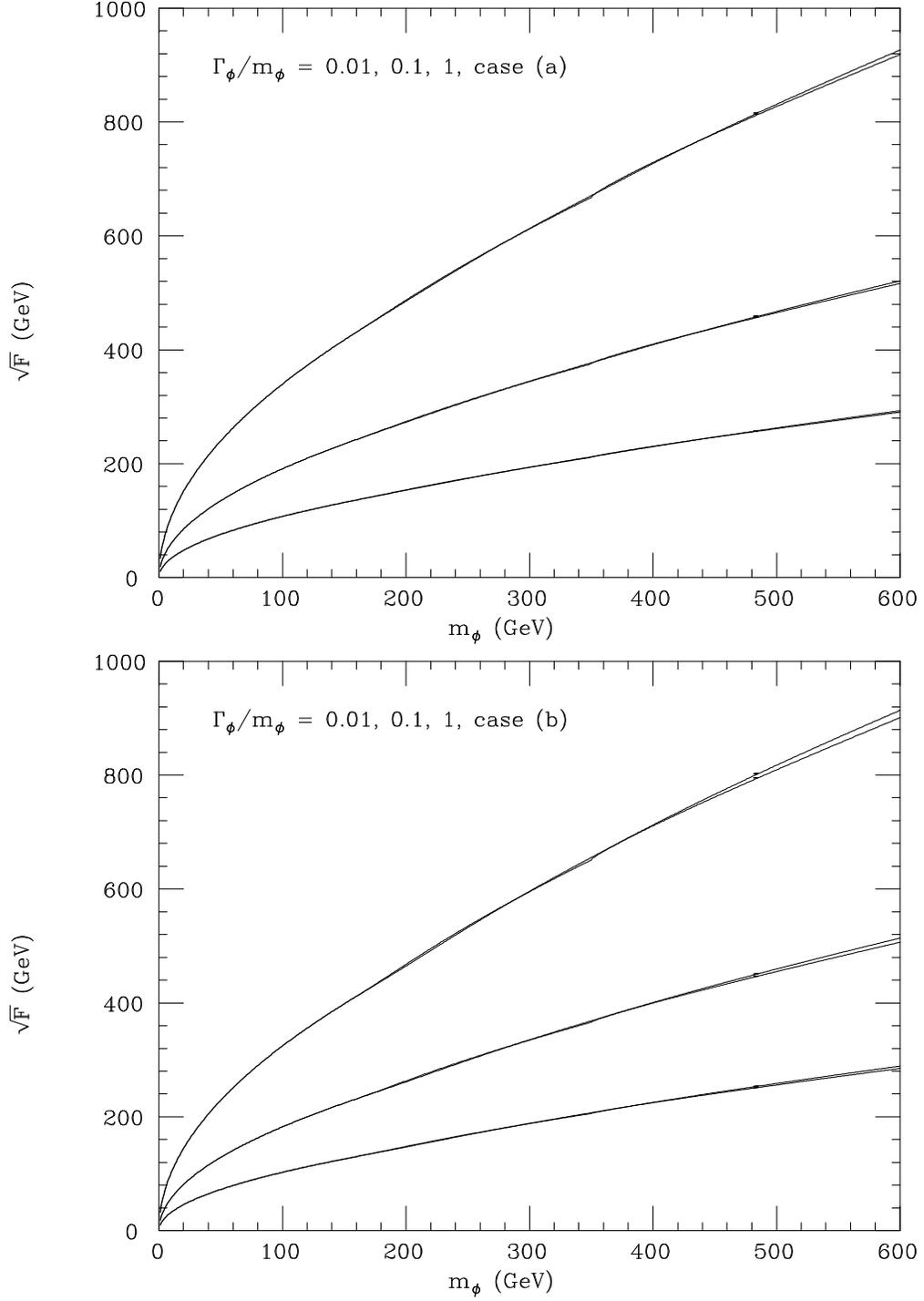

\begin{center}
\epsfig{figure=widtha.ps,height=9.5cm,angle=0}
\epsfig{figure=widthb.ps,height=9.5cm,angle=0}
\end{center}
\caption{{\it Lines corresponding to fixed values of 
$\Gamma_{\phi}/m_{\phi}$ in the $(m_{\phi},\sqrt{F})$ 
plane, for the two parameter choices of Table~1. Both 
$\phi=S$ and $\phi=P$ are considered.}}
\label{width}
\end{figure}

As already discussed in \cite{prz}, another important quantity is the 
total $\phi$ width, $\Gamma_{\phi}$, controlled by the ratios between 
the relevant mass parameters and the supersymmetry-breaking scale. 
Large values of these ratios correspond to broad, strongly coupled 
sgoldstinos: to keep the particle interpretation and the validity of 
our approximations, we must require, among the other things,
$\Gamma_{\phi}/m_{\phi} \ll 1$. To compare signal and background 
in the narrow-width approximation, we must impose more stringent 
constraints: $\Gamma_{\phi} /m_{\phi} < 10^{-1}$ when discussing decays 
into jets,  $\Gamma_{\phi} /m_{\phi} < {\rm few} \times 10^{-2}$ 
when discussing decays into photons. We then show in Fig.~\ref{width} 
contours corresponding to constant values of $\Gamma_{\phi}/m_{\phi}$ 
in the $(m_{\phi},\sqrt{F})$-plane, for the two parameter choices of 
Table~1. Since the curves for $\phi=S$ and $\phi=P$ are almost 
indistinguishable, we draw both of them simultaneously. 
As we will see in sects.~3 and 4, while the region of parameter space
of experimental interest at LEP is such that sgoldstinos can always be 
treated as very narrow resonances, this is not necessarily the case at
the Tevatron.

As a final remark on the branching ratios, we remind the reader 
that our computations have been performed at leading order. 
In analogy with the case of the SM and MSSM neutral Higgs 
bosons, we expect the NLO QCD corrections to increase the partial 
width into gluons by a factor $K \sim 1.6$ . Since the two-gluon 
decay mode is the dominant one, this may lead to a non-negligible 
suppression of the rarer decay modes. When important, this will
be taken into account in the phenomenological discussion of sect.~4.

\vspace{1cm}
\bc
{\bf 3. Production cross-sections}
\ec

The possible parton-level mechanisms for sgoldstino production at 
hadron colliders are very similar to those for the SM Higgs and 
can be easily classified:
\be
g + g \longrightarrow \phi \, ,
\label{inc}
\ee
\be
q + \ov{q} \longrightarrow V + \phi \, ,
\label{ass}
\ee
\be
q + \ov{q} \longrightarrow
q + \ov{q} + \phi \, ,
\label{fus}
\ee
\be
g + g \; {\rm or} \; q + \ov{q} 
\longrightarrow t + \ov{t} + \phi \, ,
\label{ttbar}
\ee
where $q$ denotes any quark flavour and $V$ any of the electroweak 
vector bosons, including the photon. We briefly discuss here the 
corresponding cross-sections, giving some numerical examples for 
the Tevatron. In the case of the SM Higgs, the dominant Higgs
production mechanism at the Tevatron is the one of eq.~(\ref{inc}),
with those of eqs.~(\ref{ass}) and (\ref{fus}) suppressed by roughly
one order of magnitude, and the one of eq.~(\ref{ttbar}) suppressed
by roughly an additional order of magnitude in the mass region of 
interest. An important r\^ole in these hierarchies is played by the fact 
that the SM Higgs couplings to $WW$ and $ZZ$ arise at the tree-level and 
are unsuppressed, whereas the couplings to $gg$ (and $\gamma \gamma$) 
arise at the one-loop level, and are therefore considerably suppressed. 
This compensates in part the hierarchy between strong and electroweak 
interactions, and allows for the subdominant production mechanisms to
be of phenomenological interest. In contrast with the SM Higgs, all 
sgoldstino couplings to vector boson pairs are essentially on the same 
footing. The result, which could already be guessed from our study of 
the branching ratios, is the following: the practical relevance of the 
subdominant production mechanisms at hadron colliders will be much smaller 
in the case of sgoldstinos than in the case of the SM and MSSM neutral
Higgs bosons. 

\begin{figure}[htbp]
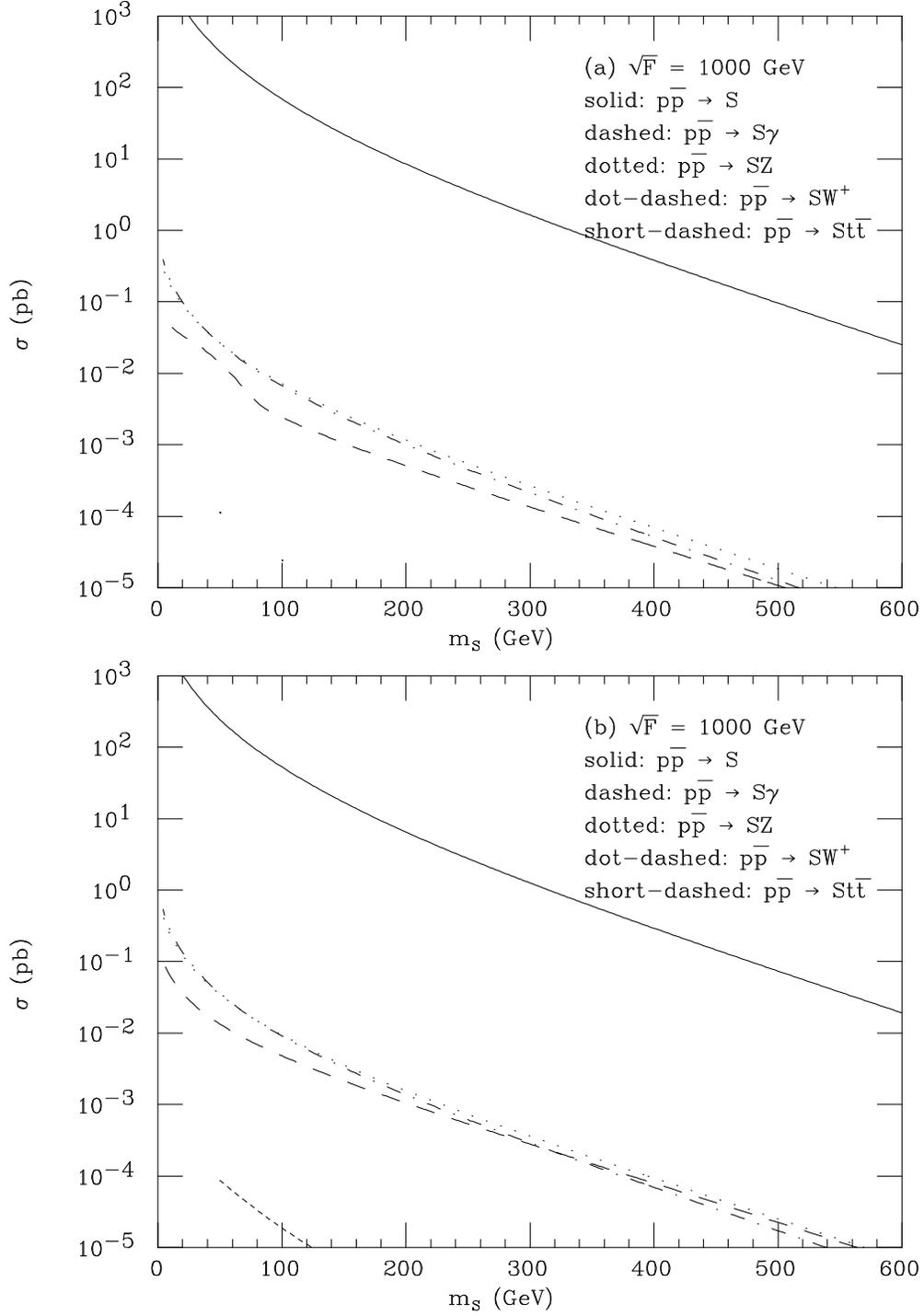

\begin{center}
\epsfig{figure=pp-tev-a.ps,height=9.5cm,angle=0}
\epsfig{figure=pp-tev-b.ps,height=9.5cm,angle=0}
\end{center}
\caption{{\it Lowest-order sgoldstino production cross-sections
at the Tevatron ($p \ov{p}$, $\sqrt{S}=2 \tev$) as functions of 
the sgoldstino mass and for $\phi=S$. Cases (a) and (b) correspond 
to the two parameter choices of Table~1.}}
\label{xstev}
\end{figure}

To produce numerical examples, we will always adopt the CTEQ5 
parametrization of the parton distribution functions \cite{cteq5} 
with $\Lambda_5=226$~MeV, corresponding to $\alpha_S(m_Z)=0.118$. 
Our results are summarized in Fig.~\ref{xstev}, which displays some of 
the lowest-order sgoldstino production cross-sections, as functions of 
the sgoldstino mass, for $p \ov{p}$ collisions at $\sqrt{S}=2 \tev$. 
The cross-sections of Fig.~\ref{xstev} have been obtained for 
$\sqrt{F}=1 \tev$. Since they are all proportional to $1/F^2$, the 
cross-sections for any other value of $\sqrt{F}$ can be obtained by 
a simple rescaling of the values of Fig.~\ref{xstev}. Cases (a) and 
(b) correspond as usual to the two representative parameter choices of 
Table~1. For simplicity, only the case of $\phi=S$ has been considered.
On the scale of Fig.~\ref{xstev}, the results for the case $\phi=P$ are
very similar.

\bc
Gluon-gluon fusion
\ec
\nopagebreak[4]
Since the sgoldstino couplings to light quarks are suppressed by 
the corresponding quark masses (as it is the case for the SM Higgs),
whilst the effective gluon-gluon-sgoldstino couplings are proportional
to $M_3/F$, the dominant production mechanism of massive sgoldstinos 
at hadron colliders is in general gluon-gluon fusion, eq.~(\ref{inc}). 
To lowest order, the partonic cross-section can be expressed in terms
of the gluonic width of the sgoldstino,
\be
\sigma ( g \, g \to \phi ) = \sigma_0 \; m_{\phi}^2 \;
\delta(s - m_{\phi}^2) \, , 
\;\;\;\;\;
\sigma_0 = {\pi^2 \over 8 m_{\phi}^3} \,
\Gamma (\phi \to g \, g) \, ,
\ee
where $\sqrt{s}$ is the total energy in the centre-of-mass frame
of the incoming partons. With the lowest-order expression for 
$\Gamma (\phi \to g \, g)$ given in \cite{prz}, we find
\be
\sigma_0 = {\pi M_3^2 \over 32 F^2} \, . 
\ee
The lowest-order proton-antiproton cross-section is then, in the 
narrow-width approximation,
\be
\sigma ( p \ov{p} \to \phi ) = 
\sigma_0 \; \tau \int_{\tau}^1 {d x \over x} f_g(x,m_{\phi}^2)
f_g(\tau/x,m_{\phi}^2) \, ,
\ee
where $\tau=m_{\phi}^2/S$, $\sqrt{S}$ is the total
centre-of-mass energy of the proton-antiproton system,
and $f_g(x,Q^2)$ is the gluon distribution function,
the same for proton and antiproton. An identical formula 
holds for the proton-proton cross-section. Notice that 
the above formulae are very similar in form to those for 
the production of a light SM Higgs boson. The only difference 
is that for the Higgs
\be
\sigma_0 = {G_F \alpha_s^2 \over 288 \sqrt{2} \pi} \, .
\ee
For a heavy Higgs, one must correct for the form factor 
originating from the top-quark loop and for the finite 
Higgs width, whilst the present approximation should 
be also applicable to heavy sgoldstinos, provided that
$m_{\phi} M_3/F \simlt 1$. The formal analogy with Higgs 
production also allows to include the NLO QCD corrections, 
by adapting the computation of \cite{spira}. For our present 
purposes, it is sufficient to work at the leading non-trivial 
order. In the phenomenological discussion of sect.~4, however,
we will include the most important QCD corrections by making 
the rough approximation $\sigma_{NLO} = K \; \sigma$, with 
$K \simeq 2$.

\bc
Associated $\phi \gamma$ production
\ec
\nopagebreak[4]
In analogy with the process $e^+ e^- \to \phi \gamma$, already 
considered in \cite{prz}, we can consider the process $q \ov{q} 
\to \phi \gamma$. This process is made possible by the existence
of effective $\gamma \gamma \phi$ and $\gamma Z \phi$ couplings,
whose explicit form can be found in \cite{prz}. They are controlled 
by the ratios $\mgg/F$ and $\mgz/F$, respectively, where $\mgg=M_1 
\cos^2 \theta_W + M_2 \sin^2 \theta_W$ and $\mgz=(M_2-M_1) \sin 
\theta_W \cos \theta_W$ . At the partonic level and before including 
QCD corrections, there are only two Feynman diagrams to compute, 
corresponding to s-channel $\gamma$ and $Z$ exchange. Neglecting 
both the quark masses and the $Z$ width, the differential 
cross-section for a given quark flavour $q$ is
\be
\label{xsg}
{d\sigma \over d \cos \theta} 
\left(q \ov{q} \rightarrow \phi \gamma \right) = 
{|\Sigma_{\phi \gamma}|^2 s \over 64 \pi N_c F^2} 
\left(1 - {m_{\phi}^2 \over s} \right)^3  
\left(1 + \cos^2 \theta \right) \, , 
\ee
where $N_c=3$ is a color factor,
\be
|\Sigma_{\phi \gamma}|^2 = { e^2  Q_q^2  \mgg^2  \over 2 s }
+
{ g_Z^2 (v_q^2 + a_q^2 ) \mgz^2  s \over 2 (s-m_Z^2)^2}
+
{e Q_q g_Z v_q \mgg \mgz \over (s-m_Z^2)} \, ,
\ee
$v_q=T_{3q}/2-Q_q \sin^2 \theta_W$, $a_q=-T_{3q}/2$, $T_{3u}=T_{3c}=
T_{3t}=-T_{3d}=-T_{3s}=-T_{3b}=1/2$, $Q_u=Q_c=Q_t=+2/3$, $Q_d=Q_s=
Q_b=-1/3$, $g_Z=e/(\sin \theta_W \cos \theta_W)$, and $\theta$ is 
the scattering angle in the centre-of-mass frame of the colliding
partons. 

\bc
Associated $\phi Z $ production
\ec
\nopagebreak[4]
Another process to be considered is $q \ov{q} \to \phi Z$,
analogous to the process $e^+ e^- \to \phi Z$ considered
in \cite{prz}. For $\phi=P$, the differential cross-section is
\be
\label{xpz}
{d\sigma \over d \cos \theta} 
\left(q \ov{q} \rightarrow P Z \right) = 
{|\Sigma_{PZ}|^2 \over 32 \pi N_c s^2 F^2} 
\sqrt{(s - m_P^2 - m_Z^2)^2 - 4 m_P^2 m_Z^2} 
\, , 
\ee
where
\be
|\Sigma_{PZ}|^2 =  
\left(
{ e^2  Q_q^2  \mgz^2  \over 2 s }
+
{ g_Z^2 (v_q^2 + a_q^2 ) \mzz^2  s \over 2 (s-m_Z^2)^2}
+
{e Q_q g_Z v_q \mgz \mzz \over (s-m_Z^2)}
\right)
\left( 
t^2 + u^2 - 2 m_P^2 m_Z^2 
\right) \, ,
\ee
and $(t,u)$ are the usual Mandelstam variables for two-body scattering. 
The cross-section for $\phi = S$ has some additional complications, 
because, as discussed in \cite{prz}, the $SZZ$ coupling has an 
additional dependence on the higgsino mass parameter $\mu_a$:
\be
\label{xsz}
{d\sigma \over d \cos \theta} 
\left(q \ov{q} \rightarrow S Z \right) = 
{|\Sigma_{SZ}|^2 \over 32 \pi N_c s^2 F^2} 
\sqrt{(s - m_S^2 - m_Z^2)^2 - 4 m_S^2 m_Z^2} 
\, , 
\ee
where
\bea
|\Sigma_{SZ}|^2 &  = &  \! 
\left[
{ e^2  Q_q^2  \mgz^2  \over 2 s }
+
{ g_Z^2 (v_q^2 \! + \! a_q^2 ) \mzz^2  s \over 2 (s-m_Z^2)^2}
+
{e Q_q g_Z v_q \mgz \mzz \over s-m_Z^2}
\right] \!
\left[ t^2 \! + \! u^2 \! + 2 m_Z^2 (2s-m_S^2) \right]
\nonumber \\ 
& + & 
{g_Z^2 \mu_a^2 m_Z^4 (v_q^2 + a_q^2 ) \over (s-m_Z^2)^2}
\left(2s -m_S^2 + {t u \over m_Z^2} \right)
\nonumber \\
& + &
{g_Z \mu_a m_Z^2 \over s - m_Z^2}
\left[ 
{g_Z (v_q^2 + a_q^2 ) \mzz \over s-m_Z^2 }
+
{e Q_q \mgz v_q \over s} 
\right] 
\left[ 2 s (s + m_Z^2 -m_S^2) \right] \, .
\eea
\newpage
\bc
Associated $\phi W$ production
\ec
\nopagebreak[4]
At hadron colliders, we can also consider the associated production
of a sgoldstino and a $W$ boson, whose partonic cross-section can 
easily be obtained from the previous ones. For $\phi=P$:
\be
\label{xpw}
{d\sigma \over d \cos \theta} 
\left(q \ov{q}' \rightarrow P W \right) = 
{|\Sigma_{PW}|^2 \over 32 \pi N_c s^2 F^2} 
\sqrt{(s - m_P^2 - m_W^2)^2 - 4 m_P^2 m_W^2} 
\, , 
\ee
where
\be
|\Sigma_{PW}|^2 =  
{ g^2 |V_{qq'}|^2 M_2^2  s \over 8 (s-m_W^2)^2}
\left( 
t^2 + u^2 - 2 m_P^2 m_W^2 
\right) \, .
\ee
As in the case of $\phi Z$ production, the cross-section for $\phi
= S$ has some additional complications, because of the additional 
dependence of the $SW^+W^-$ coupling on the higgsino mass parameter 
$\mu_a$:
\be
\label{xsw}
{d\sigma \over d \cos \theta} 
\left(q \ov{q}' \rightarrow S W \right) = 
{|\Sigma_{SW}|^2 \over 32 \pi N_c s^2 F^2} 
\sqrt{(s - m_S^2 - m_W^2)^2 - 4 m_S^2 m_W^2} 
\, , 
\ee
where
\bea
|\Sigma_{SW}|^2 &  = &   {g^2 |V_{qq'}|^2 \over 4 (s-m_W^2)^2}
\left[ {M_2^2 s \over 2}
\left[ t^2 + u^2  + 2 m_W^2 (2s-m_S^2) \right] \right.
\nonumber \\ 
& + &  \left.
\mu_a^2 m_W^4 
\left(2s -m_S^2 + {t u \over m_W^2} \right)
+
2 \mu_a m_W^2 M_2 s (s + m_W^2 -m_S^2)
\right] \, .
\eea

\bc
Vector-boson fusion
\ec
\nopagebreak[4]
The cross-sections for sgoldstino production via vector-boson fusion
can be easily calculated, starting from the effective sgoldstino
couplings to $\gamma \gamma$, $\gamma Z$, $ZZ$ and $WW$. However,
their complete analytic expressions are quite involved, and, in
analogy with the associated production, also the production via
vector boson fusion turns out to be significantly suppressed with 
respect to the production via gluon-gluon fusion. For these reasons,
we omit here a detailed discussion of this production mechanism.

\bc
Associated $t \ov{t}  \phi$ production
\ec
\nopagebreak[4]
Since the $t \ov{t} \phi$ couplings can be obtained from the Higgs
couplings of the SM (or of the MSSM) by the simple rescalings of
eqs.(\ref{subs}) and (\ref{subp}), the simplest way to obtain the
cross-sections for the associated production of a sgoldstino and 
a top-antitop pair is to take the corresponding Higgs cross-sections
in the SM (or in the MSSM) \cite{spiratop} and to rescale them by the 
appropriate factor.
\newpage
\bc
Other production mechanisms
\ec
\nopagebreak[4]
As discussed in \cite{prz}, the effective lagrangian contains some 
interaction terms that can also lead to the pair-production of a 
CP--even and a CP--odd sgoldstino, with cross-section
\be
{d \sigma \over d \cos \theta} (q \ov{q} \to S \, P ) = 
{(\tilde{m}^4_q + \tilde{m}^4_{q^c}) \over 512 \pi N_c s^2 F^4}
\left[ (s-m_S^2-m_P^2)^2-4 m_S^2 m_P^2 \right]^{3/2} \sin^2 
\theta \, ,
\ee
where $\theta$ is the scattering angle in the center-of-mass frame.
For plausible values of the parameters, we expect this cross-section
to be suppressed by the large numerical factor and the higher power 
of the supersymmetry-breaking scale at the denominator. Otherwise, 
the corresponding signal could be seen as an anomaly in the four-jet 
sample.

\vspace*{0.5cm}
To conclude this section, we comment again on the relative importance
of the different production mechanisms. We can see from Fig.~\ref{xstev}
that, for the parameter choices of Table~1, the dominant sgoldstino 
production mechanism at the Tevatron is by far gluon-gluon fusion.
The processes of eqs.~(\ref{ass}) and (\ref{fus}) are suppressed by 
roughly four orders of magnitude, and the one of eq.~(\ref{ttbar}) 
by roughly two more orders of magnitude. Therefore, we shall consider 
only inclusive signals when discussing the phenomenology at the Tevatron.

\vspace{1cm}
\bc
{\bf 4. Phenomenological discussion}
\ec

\begin{figure}[htbp]
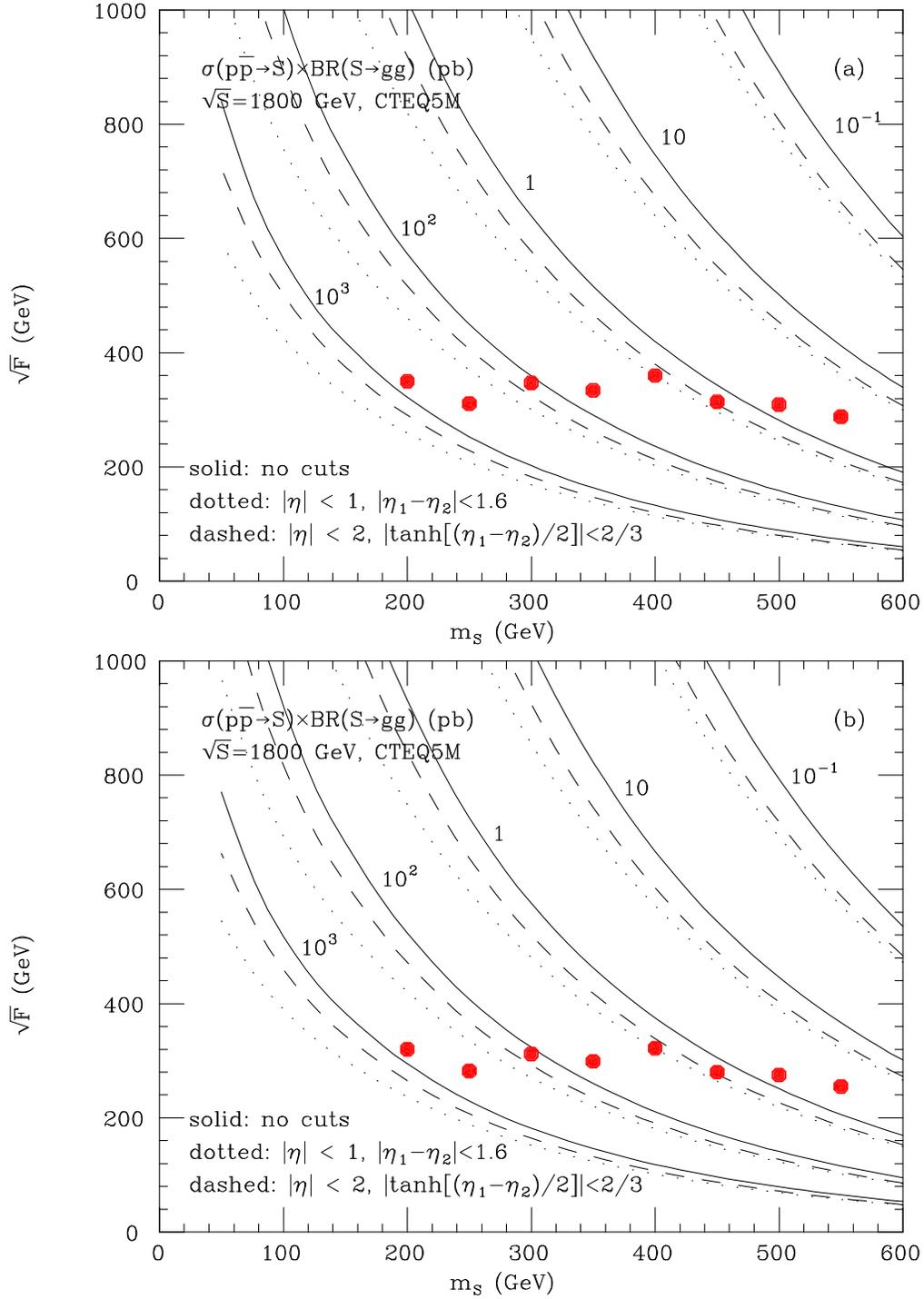

\begin{center}
\epsfig{figure=2jet-a.ps,height=9.5cm,angle=0}
\epsfig{figure=2jet-b.ps,height=9.5cm,angle=0}
\end{center}
\caption{{\it Lines corresponding to fixed values of $\sigma
(p \ov{p}  \to S + X) \times BR ( S \to g g) = 10^{-1,0,1,2,3}
\pb$, for $\sqrt{S}=1.8 \tev$ and $K=2$, in the $(m_S,\sqrt{F})$ 
plane and for the two parameter choices of Table~1. Solid lines: 
no cuts. Dashed lines: `CDF' cuts. Dotted lines: `D0' cuts. Fat 
dots: tentative estimate of the Tevatron sensitivity after run~I.}}
\label{tevgg}
\end{figure}

\bc
The inclusive di-jet signal
\ec
\nopagebreak[4]
Since the most important mechanism for sgoldstino production at the 
Tevatron is gluon-gluon fusion, and the two-gluon decay mode is the
dominant one, it is natural to consider a peak in the dijet invariant 
mass distribution as a possible signal to be looked for. As a first 
element to measure the Tevatron sensitivity, we draw in 
Fig.~\ref{tevgg} contours of constant $\sigma(p \ov{p} \to \phi + X) 
\times BR (\phi \to g g)$ in the $(m_{\phi},\sqrt{F})$ plane, for 
$\sqrt{S}=1.8 \tev$ and the two parameter choices of Table~1. Since 
the curves for $\phi=S$ and $\phi=P$ are almost indistinguishable, we 
consider only the case $\phi=S$. To account approximately for the NLO 
QCD corrections \cite{spira}, we have multiplied the LO result by a 
K--factor $K=2$.  

To interpret the curves of Fig.~\ref{tevgg}, we need some additional 
information on the SM backgrounds and on the experimental sensitivity. 
Fortunately, we can rely on two recent analysis by CDF \cite{cdfdijet} 
and D0 \cite{d0dijet}, devoted to the search of new particles decaying to 
dijets, and thus applicable to sgoldstinos. Using $106 \pb^{-1}$ of data 
collected at $\sqrt{S}=1.8 \tev$, and requiring that both jets have 
pseudorapidity $|\eta|<2.0$ and a scattering angle in the dijet 
center-of-mass frame $|\cos \theta^*| \equiv |\tanh [ (\eta_1 - 
\eta_2)/2]|<2/3$, CDF plots and tabulates the $95 \%$ c.l. upper 
limit on the cross-section times branching ratios for narrow 
resonances ($\Gamma/M < 0.1$) decaying into dijets. D0 uses
$104 \pb^{-1}$, requires $|\eta|<1.0$ and $|\eta_1 - \eta_2|<1.6$,
and plots a  $95 \%$ c.l. upper limit on the cross-section times 
branching ratio times acceptance for three representative models of
new physics. We have used the published CDF and D0 data and our 
calculation of the sgoldstino cross-sections and branching ratios 
to draw in Fig.~\ref{tevgg} a sequence of fat dots, representing our 
tentative estimate of the Tevatron sensitivity after run~I. This line 
was obtained by selecting, for each value of the sgoldstino mass (from 
200 to 600~GeV, in steps of 50~GeV), the more stringent of the CDF 
and D0 limits on $\sqrt{F}$, obtained under the assumption $\Gamma_\phi
/m_\phi < 0.1$. By comparing with Fig.~\ref{width}, however, we can see 
that for increasing sgoldstino masses this assumption is more and more 
strongly violated. As a result, we expect to have made a stronger and 
stronger overestimate of the present bounds for increasing mass values 
in the region above 300 GeV. 

To have an idea of the Tevatron sensitivity after run~II, for each given 
value of the sgoldstino mass we can rescale the corresponding value of 
$\sqrt{F}$ by a factor $20^{1/8} \simeq 1.5$. This amounts to making the 
na\"{\i}ve assumptions that the cross-section does not vary much when
changing $\sqrt{S}$ from $1.8$ to $2.0$~TeV, and that the error in the 
cross--section measurement will scale as $1/\sqrt{L}$, where $L$ is the 
integrated luminosity.

\bc
The inclusive di-photon signal
\ec
\nopagebreak[4]
In the case of the SM and MSSM neutral Higgs bosons, the two-photon decay
mode has a very suppressed branching ratio, at most ${\cal O}(10^{-3})$
in the mass region between 100 and 150~GeV, and much smaller for larger 
masses. As a result, the diphoton signal is marginal for Higgs searches
at the Tevatron, and it is much more convenient to exploit other decay 
modes, in conjunction with associated production mechanisms. In contrast,
sgoldstinos have diphoton branching ratios well above ${\cal O} (10^{-2})$
over the whole mass range between a few and 1000 GeV, and possibly larger 
inclusive production cross-sections, so we can expect the inclusive diphoton 
signal to play a major r\^ole in sgoldstino searches at the Tevatron.

\begin{figure}[htbp]
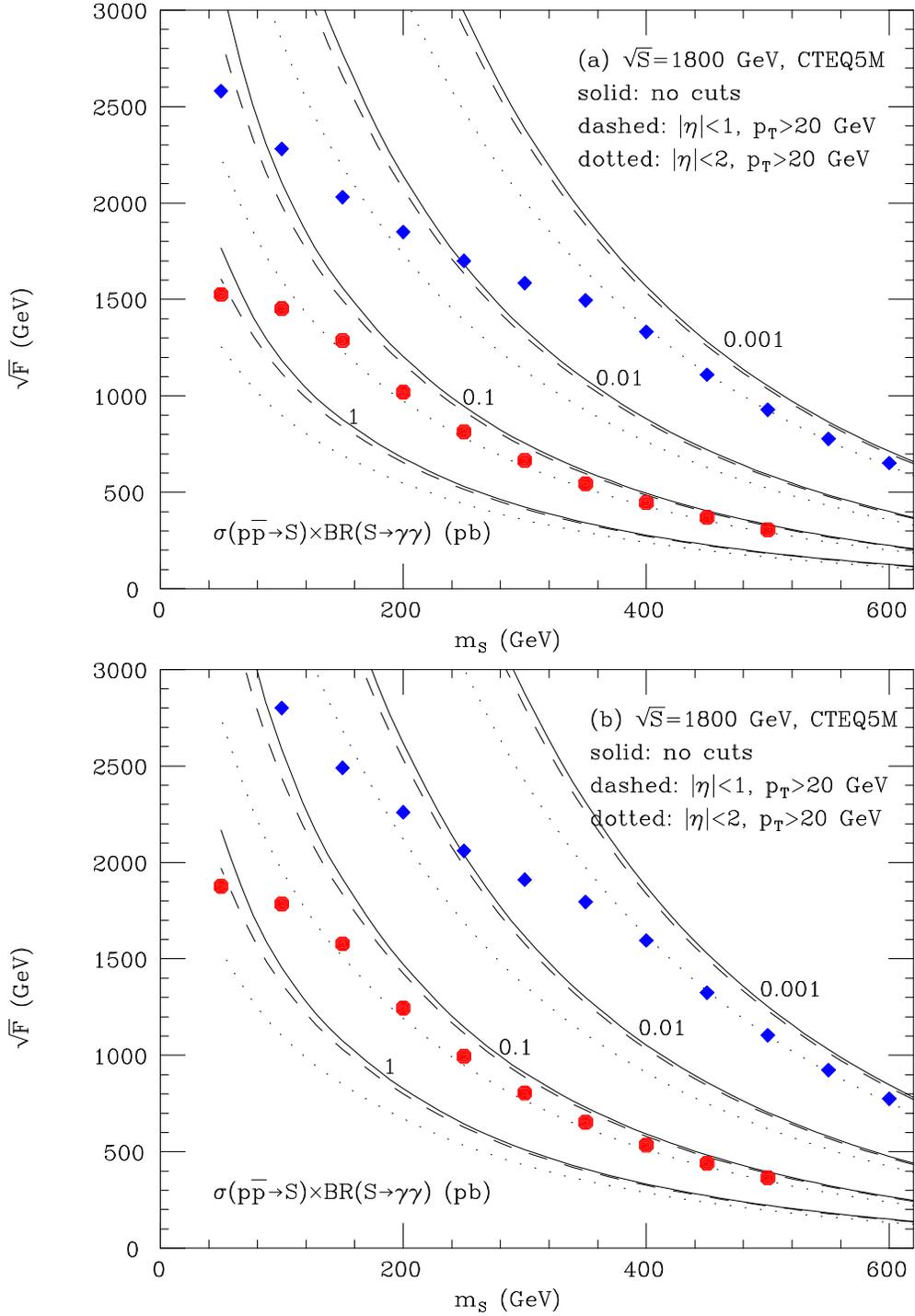

\begin{center}
\epsfig{figure=2gamma-a.ps,height=9.5cm,angle=0}
\epsfig{figure=2gamma-b.ps,height=9.5cm,angle=0}
\end{center}
\caption{{\it Lines corresponding to fixed values of $\sigma
(p \ov{p}  \to S + X) \times BR ( S \to \gamma \gamma) = 
10^{-3,-2,-1,0}  \pb$, for $\sqrt{S}=1.8 \tev$ and $K=1$, in 
the $(m_S,\sqrt{F})$ plane and for the two parameter choices 
of Table~1. The fat dots and diamonds represent our tentative 
estimates of the Tevatron sensitivity after run~I and after 
run~II, respectively.}}
\label{tevff}
\end{figure}

As a first piece of information, we display in Fig.~\ref{tevff} contours of 
constant $\sigma(p \ov{p} \to \phi + X) \times BR (\phi \to \gamma \gamma)$
in the $(m_{\phi},\sqrt{F})$ plane, for $\sqrt{S}=1.8 \tev$ and the two 
parameter choices of Table~1. Since the curves for $\phi=S$ and $\phi=P$ 
are almost indistinguishable, we consider only the case $\phi=S$. In this
case, the NLO QCD corrections \cite{spira} are not included, since they have
two competing effects that approximately cancel: an increase in the inclusive
production cross-section, but also a similar increase in the two-gluon 
partial width, which correspondingly reduces the two-photon branching ratio.

To interpret the curves of Fig.~\ref{tevff}, we need some additional 
information on the SM backgrounds and on the experimental sensitivity.
Again, we can rely on some recent analysis by CDF \cite{cdfdiph} and 
D0 \cite{d0diph}, as well as on some useful information contained
in a recent study for run~II \cite{lmdiph}. The CDF study uses $100 
\pb^{-1}$ of data collected at $\sqrt{S}=1.8 \tev$, and requires that 
both photons have pseudorapidity $|\eta|<1$ and transverse energy
$E_T>22 \gev$. Since the study looks specifically at the invariant 
mass distribution of diphotons above 50~GeV, it is relatively simple
to translate its results into a tentative sensitivity curve in
the $(m_{\phi},\sqrt{F})$ plane. The D0 studies were performed with
slightly different goals, and some work would be required in order to 
optimize their data analysis for setting limits on sgoldstinos. On
the contrary, the study in \cite{lmdiph}, performed with the cuts 
$|\eta|<2$ and $E_T>20 \gev$, comes with an analytical formula 
for the background and can be easily used to obtain a first, rough 
estimate of the run~II sensitivity. For example, following \cite{lmdiph},
we can assume a diphoton identification efficiency $\epsilon=0.8$ and
combine the finite sgoldstino width and the experimental resolution in 
the diphoton invariant mass by defining a quantity $[\Delta/(1 \gev)] =
\sqrt{ [\Gamma_{\phi}/(1 \gev)]^2 + (0.35)^2 \, [m_{\phi}/(1 \gev)]}$.
We can then ask that, if $S$ is the number of signal events and $B$
the number of background events in a window of width $1.2 \, \Delta$
centered around $m_{\phi}$, either $S/\sqrt{B}>5$ (if $B>1$) or
$S>5$ (if $B<1$). Doing so, and leaving a more detailed and reliable
analysis to our experimental colleagues, we could draw in Fig.~\ref{tevff} 
two series of fat dots and diamonds, representing our tentative estimates 
of the Tevatron sensitivities after run~I and after run~II, respectively. 
A more sophisticated study could proceed along the lines of \cite{cl} 
and make full use of the double differential cross-section in the diphoton 
invariant mass and in the scattering angle, to take into better account 
the finite-width effects (important for large sgoldstino masses) and to
reduce the dependence on the cuts (important for small sgoldstino masses). 
For our present purposes, however, it is sufficient to observe that, for 
the two parameter choices of Table~1, searching for the diphoton signal 
is by far the most powerful way of constraining the sgoldstino parameter 
space.~\footnote{The range of accessible sgoldstino masses may extend
beyond the right border of Fig.~\ref{tevff}. To reliably explore that 
region, however, we should take into account the effects of sgoldstino
decays into neutralinos, charginos and eventually gluinos.}

Whilst the searches for very massive sgoldstinos should be relatively 
straightforward, an interesting phenomenological question is how to 
extend and maximize the Tevatron reach in the region of small masses,
$m_{\phi} \ll 100 \gev$. In this respect, it may be useful to relax as 
much as possible the trigger and selection requirements on the photon
transverse energy. Also, in the region of very small masses, the
associated production mechanisms with electroweak gauge bosons may
play a useful role. 

\vspace*{1.0cm}
{\bf Acknowledgements. }
We would like to thank A.~Brignole, A.~Castro, G.~Landsberg, M.L.~Mangano, 
M.~Spira, A.S.~Turcot and D.~Wood for useful discussions and suggestions.
%
%\newpage
%

%
\end{document}